\documentclass[pdftex,showpacs,superscriptaddress]{revtex4}

\pdfoutput=1
\usepackage{graphicx}
\usepackage{dcolumn}
\usepackage{bm}
\usepackage{amssymb}
\usepackage{amsmath}

\begin{document}

\title{Bow wave from ultra-intense electromagnetic pulses in plasmas}

\author{T. Zh. Esirkepov}
\affiliation{Kansai Photon Science Institute, JAEA, Umemidai 8-1, Kizugawa, Kyoto 619-0215, Japan}
\author{Y. Kato}
\affiliation{The Graduate School for the Creation of New Phototnics Industries, Kurematsucho 1955-1, Nishiku, Hamamatsu, Shizuoka 431-1202, Japan}
\author{S. V. Bulanov}
\altaffiliation[Also at ]{A. M. Prokhorov Institute of General Physics of RAS, Moscow, Russia.}
\affiliation{Kansai Photon Science Institute, JAEA, Umemidai 8-1, Kizugawa, Kyoto 619-0215, Japan}

\begin{abstract}

We show a new effect of the bow wave excitation by
an intense short laser pulse propagating in underdense plasma.
Due to spreading of the laser pulse energy in transverse direction,
the bow wave causes a large-scale transverse modulation of the electron density.
This can significantly increase the electric potential of the wake wave
since the wake wave is generated in the region much wider than the laser pulse waist.
\end{abstract}

\maketitle


Bow waves are observed at various conditions in fluids, gases and plasmas.
A bow wave is formed in the water at the bow of a moving ship \cite{bib:Ship}
in addition to Kelvin waves left behind;
it defines the outer boundary of the ship wakes.
Reducing this effect is an important aim of shipbuilding,
since bow waves take away a portion of the ship energy
and can damage shore facilities.
In air, a spectacular bow wave
accompanied with a vapour cone
manifesting the Prandtl-Glauert singularity \cite{bib:Prandtl-Glauert}
is excited by an aircraft when it breaks the sound barrier.
In space, plasma bow waves appear at various scales.
A bow shock is formed where the Earth's magnetosphere
is squeezed by the solar wind flow \cite{bib:Solar-wind}.
The largest ever seen bow waves were observed in the collision of galaxies
by the Spitzer Space Telescope \cite{bib:Spitzer}.
As these examples show,
bow waves lead to a vast range of 
physical processes such as
a transverse transfer of momentum and matter
and particle acceleration.

As is well known,
an intense short laser pulse
propagating in underdense plasma
excites wake waves, similarly to a moving ship producing the Kelvin waves.
This analogy underlying a fruitful exchange of ideas between hydrodynamics and plasma physics
can be extended further.
In this Letter we demonstrate that
a laser pulse can excite a bow wave in collisionless plasma.
Like a bow wave from a ship,
the bow wave from a laser pulse
spread aside the energy and the momentum of the laser pulse.
Although generation of shock waves in laser plasmas
has been known shortly after the first laser was built
\cite{bib:Koopman},
to our best knowledge,
the bow waves formed by relativistic electrons
at the ultra-short laser pulse propagating in plasma
have not been discussed so far. 

The laser-driven wake waves in plasmas
are used in the laser wake field accelerator (LWFA)
\cite{bib:LWFA} and the flying mirror (FM) \cite{bib:FM} concepts.
A tightly focused and sufficiently intense laser pulse
excites a wake field \cite{bib:LWFA}
in the so-called blowout regime
\cite{bib:QME-bunch,bib:Blow}
or bubble regime \cite{bib:Bubble},
where,
in addition to a longitudinal push,
the laser pulse expells electrons also
in transverse direction,
forming a cavity void of electrons
in the first period of the wake wave.
Thus a finite waist of the laser pulse
results in a transverse motion of electrons,
which leads to a transverse wave-breaking \cite{bib:TWB}
underlying one of the mechanisms of self-injection
of electrons into the accelerating phase of the wake wave \cite{bib:QME-bunch}.
We show that
when the waist of a sufficiently intense laser pulse
is less than the wake wave length,
the expelled electrons can not be confined near the cavity
and their transverse outflow forms a bow wave.
The bow wave should be distinctly seen in the limit of
high intensity and narrow waist laser pulses
and low-density plasma,
which correspond to the conditions for
generating higher energy electrons by laser wakefield acceleraion
\cite{bib:QME-bunch,bib:Leemans-GeV}.

\begin{figure}[ht]
\includegraphics{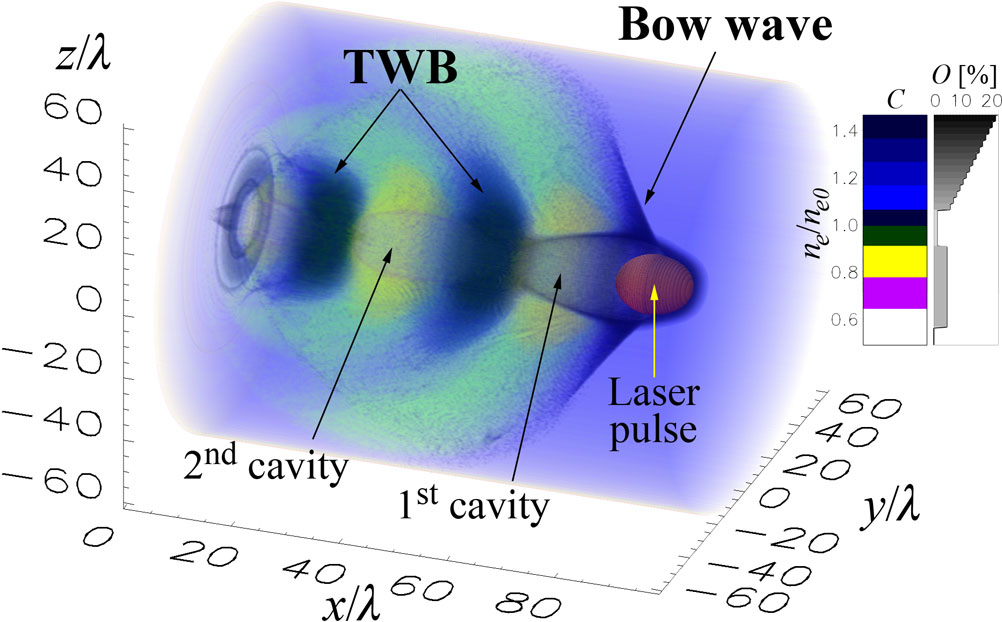}
\caption{\label{fig:1}
(color).
Electron density
(coded into color, $C$, and opacity, $O$, with ray-tracing)
in the wake of the laser pulse,
which is represented by the electromagnetic energy density
isosurface $(E^2+B^2)(m_e\omega c/e)^2=4$ (red),
at $t = 100\times 2\pi/\omega$.
}
\end{figure}

\begin{figure}[ht]
\includegraphics[scale=0.9]{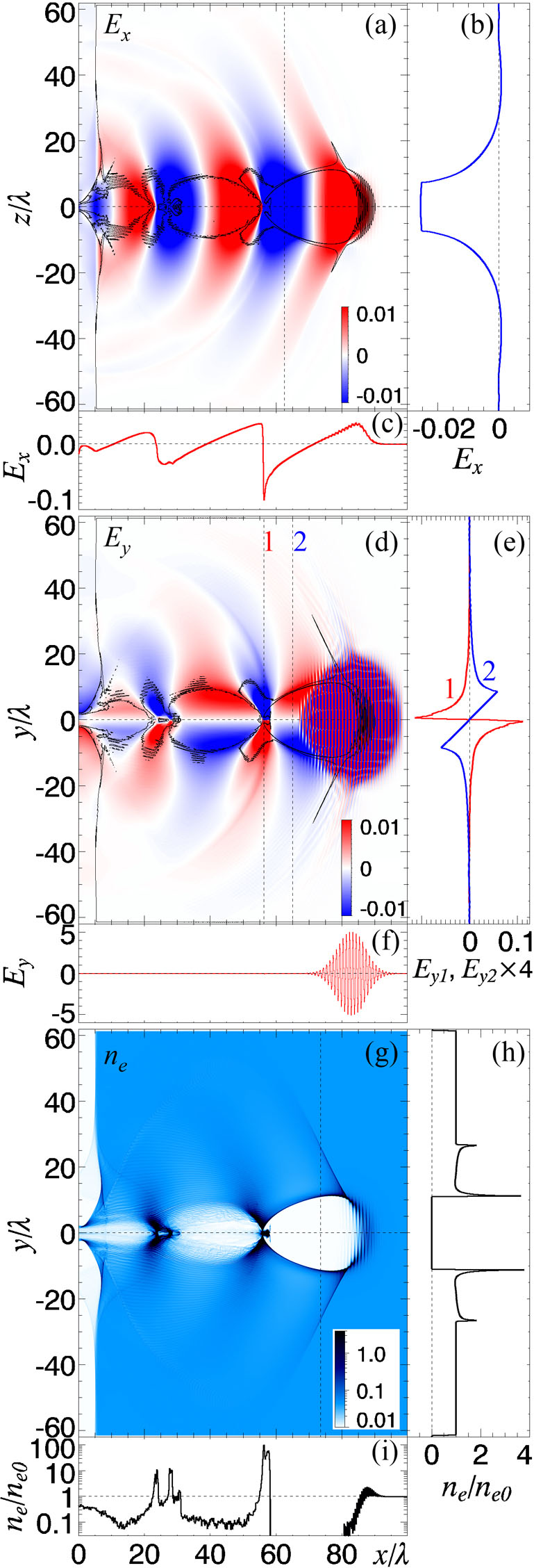}
\caption{\label{fig:2}
(color).
2D cross-sections of the electric field components
$E_x$ (a) and $E_y$ (d) and of the electron density (g)
at $t = 100\times 2\pi/\omega$.
Vertical (b,e,h) and horizontal (c,f,i) 1D cross sections
of the E-field and density
correspond to the positions shown by the
dashed lines. In (a,d) the electron density
curves are drawn by the black curves for $n_e/n_{e0} = 0.75$, $1.5$.
}
\end{figure}


We present the results of three-dimensional (3D)
particle-in-cell (PIC) simulations with
the Relativistic Electro-Magnetic Particle-mesh
(REMP) code \cite{bib:Dens-dec}
performed on Altix 3700 supercomputer at JAEA-Tokai.
The laser pulse
with the initial intensity $I=6\times 10^{19}$W/cm$^2\times(1\mu$m$/\lambda)^2$,
corresponding to the dimensionless amplitude $a=eE_0/\omega m_e c = 6.62$,
propagates along the $x$-axis;
it is linearly polarized in the direction of the $y$-axis
and has the size of
$\ell_{x}\times \ell_{y}\times \ell_{z}=10\lambda\times 10\lambda\times 10\lambda$.
The pulse is incident on a fully ionized plasma slab
with the electron density
$n_e = 1.14\times 10^{18}$cm$^{-3}\times(1\mu$m$/\lambda)^2$.
Here $\lambda$ and $\omega$ are the laser wavelength and frequency,
$e$ and $m_e$ are the charge and mass of electron,
$E_0$ is the magnitude of the laser pulse electric field,
and $c$ is the speed of light in vacuum.
We consider the electron bow wave,
when the ion response can be neglected,
i.~e. the ion-to-electron mass ratio is $m_i/m_e\rightarrow\infty$.
The simulation grid dimensions are
$4000\times 992\times 992$ along $x$, $y$ and $z$ axes;
the grid mesh sizes are $dx=\lambda/32$, $dy=dz=\lambda/8$;
total number of quasi-particles is $2.3\times 10^{10}$.
For observing the bow wave,
the number of quasi-particles must be sufficiently high
since the density of quasi-particles expelled in transverse direction
drops with a distance $r$ from the axis as $r^{-2}$ (in 3D configuration).
In the figures, the time and spatial units are the laser wave period
and wavelength, respectively.

The laser pulse excites a strong wake wave,
whose first period forms a cavity completely void of electrons,
Figs.~\ref{fig:1}, \ref{fig:2}.
In addition to a longitudinal push,
the laser pulse drives electrons also
in transverse direction, forming the bow wave.
As seen in Fig.~\ref{fig:2}(a),
the longitudinal electric field ($E_x$)
is formed in a region which is much greater
in transverse direction than the
cavity width corresponding to the first period of the wake wave.
Due to this, the electric potential, $\varphi$, 
is not limited by the transverse size of the cavity.
We note that at present conditions
the length of the cavity (i.~e., its size along $x$-axis)
is 1.5 times 
greater than its width (the size in transverse direction).
In the regimes not favourable for the bow wave formation,
e.~g. when the bow wave does not detach from the cavity,
the electric potential in the cavity
is determined by the smallest dimension of the cavity,
either longitudinal or transverse.

A strong transverse wave-breaking (TWB)~\cite{bib:TWB}
forms the transverse outflow of electrons,
which increases the electric potential in the subsequent periods
of the wake wave, similarly to the bow wave.
However, in contrast to the bow wave, which arises because
electrons are directly pushed by ponderomotive force of the laser pulse,
the TWB wave is due to an electrostatic field of the wake wave itself
caused by the finite waist of the laser pulse.

\begin{figure}
\includegraphics{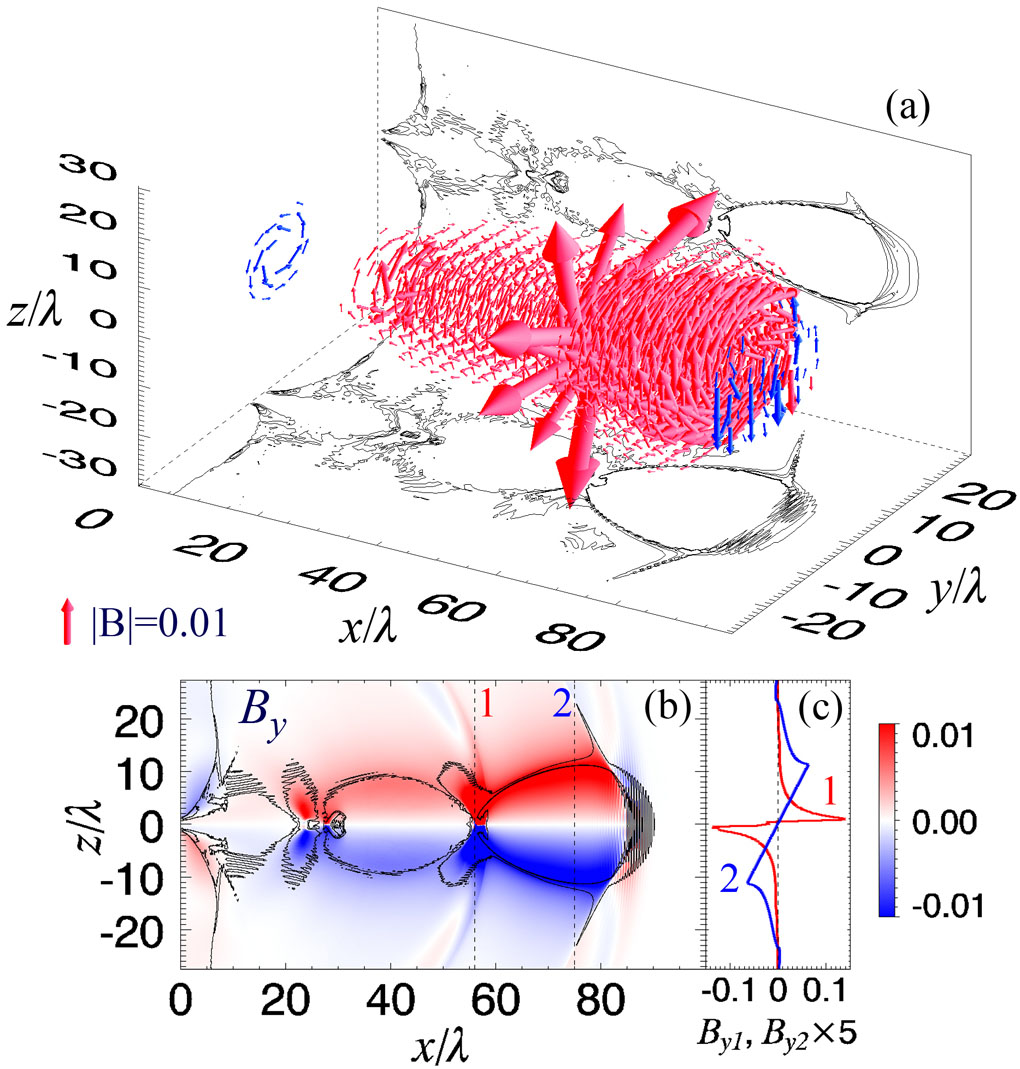}
\caption{\label{fig:3}
(color).
(a) Magnetic field, $B$, around the cavities;
laser frequency is filtered out.
Red arrows for clockwise, blue -- for counterclockwise vector $B$.
On side panels -- 2D cross-sections of the electron density
curves for $n_e/n_{e0} = 0.8,1.4,2,2.6$.
(b,c) 2D and 1D cross section of $B_y$ with electron density curves
as in Fig.~\ref{fig:2}.
Both frames for $t = 100\times 2\pi/\omega$.
}
\end{figure}

The electron motion in the wake wave, bow wave
as well as motion due to TWB, forms a transient electron density modulations
with associated electromagnetic field.
Fig.~\ref{fig:3} shows the magnetic field co-moving
with the cavities. The polarity of the magnetic field corresponds
to the forward motion of a negatively charged object.
The strong magnetic field associated with the
region of the electron density compression due to TWB
can be described in the approximation of a Lienard-Wiechert potential
of a moving point charge~\cite{LW-charge}.

\begin{figure}
\includegraphics{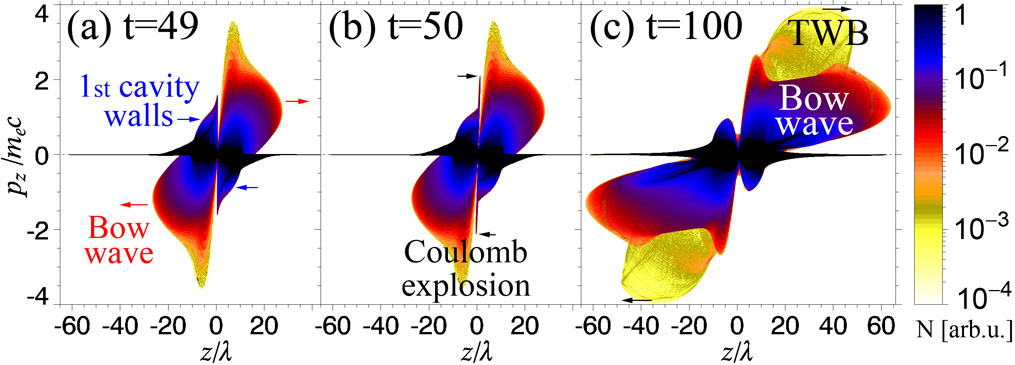}
\caption{\label{fig:4}
(color).
Electron phase space projection onto $(z,p_z)$ plane.
}
\end{figure}

As shown in the electron phase plot, Fig.~\ref{fig:4}(c),
the transverse momentum of the electrons undergone TWB
behind the first cavity
is higher than that of electrons in the bow wave.
This is due to Coulomb explosion of the electron flow
concentrated near the axis, Fig.~\ref{fig:4}(a,b).



The shape of the cavity in the electron density behind the laser pulse
can be deduced from the fact that,
in the limit of a well-developed bow wave,
all the electrons on a way of the laser pulse are pushed aside
forming a multiflow motion owing to the collisionless nature of the low density plasma.
The region void of electrons is positively charged
and attracts the unpertubed electrons from aside.
The transverse (radial) component of the electric field
near the axis is proportional to the radius, $E_\perp=2\pi n_e e r$.
It is easy to show that for the laser pulse waist smaller than $2\pi c a/\omega_{pe}$,
where $\omega_{pe} = 4\pi n_e e^2/m_e)^{1/2}$ is the Langmuir frequency,
the bulk electrons remain nonrelativistic.
The electrons start to perform an oscillation
with frequency $\Omega\propto\omega_{pe}$,
determined by the geometry of the cavity.
After one quarter of the oscillation period,
$\pi/2\Omega$, they reach the cavity axis.
Therefore the characteristis shape of
the cavity formed in the first wake wave period
can be approximated by the curve 
\begin{equation}
r = R_c \cos\left( \Omega(t-x/v_{ph}) \right)
\end{equation}
for $-\pi/2 < \Omega(t-x/v_{ph}) < 0$,
where $v_{ph}$ is the phase velocity of the wake wave
equal to the group velocity, $v_g$, of the laser pulse.

Comparing the kinetic energy of the expelled electrons, $\approx a m_e c^2$,
with their potential energy at the cavity periphery,
$e\varphi = \pi n_e e^2 R_c^2$, where $R_c$ is the cavity radius,
we find a condition of the bow wave formation:
$R_c<2c{a}^{1/2}/\omega_{pe}=(\lambda/\pi)({an_{cr}/n_e})^{1/2}$.
If we assume that the cavity transverse size is not greater than the laser pulse waist,
$D_{las}\ge R_c$,
then the condition of the bow wave formation can be rewritten
in a stronger form as $D_{las}\leq 4c{a}^{1/2}/\omega_{pe}$.
In the opposite limit, $D_{las}\gg 4c{a}^{1/2}/\omega_{pe}$,
the momentum acquired by electrons from the laser pulse
is not enough for escaping in transverse direction, so
the bow wave is closely attached to the cavity boundary.

As is well known,
in one-dimensional (1D) case the laser pulse energy depletion
length is given by $l_{dep}\sim l_{las}(\omega/\omega_{pe})^2$,
where $l_{las}$ is the laser pulse length,
which comes from the balance of the laser energy and the energy left in the wake wave
behind the pulse, when it is assumed that the electron energy is
of the order of $a^2m_ec^2/2$ \cite{bib:Depletion}.
In the 3D configuration,
when the laser pulse is tightly focused, $D_{las}\leq 4c{a}^{1/2}/\omega_{pe}$,
so that the bow wave is generated,
the kinetic energy of electrons being expelled in transverse direction
is $\approx am_ec^2$,
which results in the estimation for the depletion length:
$l_{dep}\sim a l_{las}(\omega/\omega_{pe})^2$,
i.~e. by the factor $a$ larger than in the 1D case.

In the LWFA concept \cite{bib:LWFA}
the energy of the electron accelerated by the wake field
scales as
\begin{equation}
{\cal E} \approx 2 e \Delta\varphi \gamma_{ph}^2
,
\end{equation}
where $\Delta\varphi=\varphi_{max}-\varphi_{min}$
is the potential difference in the wake field
and $\gamma_{ph}=(1-v_{ph}^2/c^2)^{-1/2}$ is the Lorentz factor corresponding
to the wake wave phase velocity.
In the 1D configuration
the electromagnetic wave propagates with the group velocity, $v_g$,
corresponding to 
$\gamma_{g} =(1-v_{g}^2/c^2)^{-1/2} = \gamma_{ph} \simeq a^{1/2}\omega/\omega_{pe}$
in the limit $a\gg 1$.
For the optimal conditions of the acceleration,
$e\Delta\varphi/m_e c^2 \simeq a^2/2$,
thus we obtain \cite{bib:Scaling-I}
\begin{equation}
{\cal E}^{(1D)} \approx a^3 \frac{\omega^2}{\omega_{pe}^2} m_e c^2
.
\end{equation}
Multi-dimensional effects modify the scaling law of the electron acceleration.
In the cavity, formed behind a tighly focused laser pulse,
close to the threshold of the bow wave formation,
the electrostatic potential is mainly determined by the cavity transverse size,
$\Delta\varphi \approx \varphi \simeq \pi n_e e D_{las}^2/4$.
For a narrow laser pulse its group velocity
becomes strongly dependent on its waist, $\gamma_{ph} \simeq D_{las}/\lambda$.
The laser pulse waist, in its turn, depends on the pulse evolution,
i.~e. self-focusing and electron density cavity formation.
We find that the energy of electrons accelerated in the wake wave
generated by a sufficiently strong,
$a \gtrsim e\varphi/m_e c^2 = (\pi/2)^2 (D_{las}/\lambda)^2 (\omega_{pe}/\omega)^2$,
and narrow,
$D_{las}\lesssim 4c{a}^{1/2}/\omega_{pe}$,
laser pulse,
does not depend on the laser pulse amplitude \cite{bib:N.epoch}
\begin{equation}
{\cal E}^{(3D)} \approx \frac{\pi^2}{2} \frac{D_{las}^4}{\lambda^4} \frac{\omega_{pe}^2}{\omega^2} m_e c^2
.
\end{equation}
We note that here the condition for $a$ implies
${\cal E}^{(3D)} \lesssim 8 \pi^{-2} a^2 (\omega/\omega_{pe})^2 m_e c^2$.

The shape of an outer boundary of the bow wave
can be determined in the limit of a small density of electrons forming the bow wave,
when these electrons can be treated as test particles.
In this approximation,
particles acquire initial momentum near the axis
and then move in a potential of a positively charged
semi-infinite wire of the cavity radius, $R_c$, propagating with velocity $v_{ph}$.
This motion is described by the Hamiltonian 
$H 
=\left(m_e^2c^4+p_x^2c^2+p_y^2c^2\right)^{1/2}
-v_{ph} p_x
+ e\sigma \ln\left[\xi+\left(\xi^2+y^2\right)^{1/2}\right] $,
for $y\ge R_c$.
On the particle trajectory in phase space
the Hamiltonian is equal to
$H_{0} =
\left(m_e^2c^4+p_{x0}^2c^2+p_{y0}^2c^2\right)^{1/2}
-v_{ph} p_{x0}
+ e\sigma \ln\left[y_0\right]$,
where $\xi = x-v_{ph}t$,
$\sigma \approx \pi n_e e R_c^2$
is the linear density of the wire,
$p_{x0}$ and $p_{y0} $ 
are the initial momentum components at $y_0 = R_c$.
Neglecting the longitudinal momentum
and assuming $p_{y0} = a m_e c$,
we obtain
$y_{max} \simeq R_c \exp\left[ (1+a^2)^{1/2}(2c/R_c\omega_{pe})^2 \right]$.
Near the axis the outer boundary of the bow wave
can be approximated by the straight line
$v_{\perp 0}(x-v_g t) + (v_g-v_{\parallel 0})y = 0$,
where $v_{\parallel 0}$ and $v_{\perp 0}$
are the components of the initial velocity of expelled electrons.
The inclination of the line is determined by the expression
$\tan\theta=dy/dx=v_{\perp 0}/(v_g-v_{\parallel 0})$.
Extrapolating the approximation of the outer boundary of the bow wave
by a line over the distance equal to the half-length of the cavity, $\ell_c/2$,
we can estimate an addition to the cavity potential
due to the bow-wave:
$\varphi^{(Bw)}\approx \sigma (1+\ln[\ell_c \tan\theta/2R_c])$.


Among various effects associated with the bow wave we notice the following.
In the multi-species underdense plasma,
on the ion time scale the bow wave structure becomes enriched by contribution
from the ion dynamics.
The bow wave has a complex structure,
where modulations of lighter components develop faster,
determining the collisionless shock wave structure~\cite{bib:Koopman,bib:colshock}.
The ion component of the bow wave is accelerated
due to the laser ponderomotive potential
and due to moving potential of the electron component of the bow wave.
Transverse electric currents, associated with the bow wave
from a laser pulse with strongly squeezed focal spot,
should form large-scale quasi-static magnetic field
and, due to instabilities,
eventually should take a shape of separated jets.
A portion of the transition radiation
emitted at the plasma boundary interface
\cite{bib:Leemans-TR}
can be attributed to
a change of the electric charge density,
associated with the bow wave propagation,
especially in a  non-uniform plasma.


In conclusion,
a tightly focused intense laser pulse propagating in underdense plasma,
in addition to wake wave generation,
excites bow waves and TWB waves,
which spread the laser pulse energy in transverse direction.
The effects of the bow and TWB waves excitation
are essentially multidimensional.
The bow wave increases the
electric potential of the wake wave
in comparison with the regimes where the bow wave can not detach from the cavity.
The bow wave facilitates the transverse wave-breaking \cite{bib:TWB}
which causes the self-injection of electrons into the accelerating phase of the wake field.
In its turn, the wave-breaking forms a transverse flow of
electrons similar to the electron motion in the bow wave.

Recent advances in LWFA \cite{bib:QME-bunch,bib:Leemans-GeV}
reveal a tendency
to use greater laser intensity and lower plasma density,
aiming at higher energy of fast electrons.
This makes the bow wave excitation inevitable.
In such the regimes the multi-dimensional effects
should modify
the scaling laws of LWFA acceleration
based 
on a one-dimensional approximation
\cite{bib:LWFA,bib:Blow,bib:Scaling-I}
or on the assumption that the cavity size is determined by the laser pulse amplitude
\cite{bib:Scaling-II}.

\begin{acknowledgements}
This work is supported by the
Grant-in-Aid for Scientific Research (A), 20244065, 2008
from MEXT (Japan) and VITP-ELI (France).
\end{acknowledgements}

\end{document}